\documentclass[prl,amsmath,amssymb,twocolumn]{revtex4}
\usepackage{graphicx}
\usepackage{color}
\usepackage{amsmath}
\usepackage{amssymb}

\usepackage{bm}
\usepackage{bbm}

\newcommand{\CH}{\mathbb{C}\otimes\mathbb{H}}

\newcommand{\RCHO}{\mathbb{R}\otimes\mathbb{C}\otimes\mathbb{H}\otimes\mathbb{O}}

\begin{document}

\title{Towards a unified theory of ideals}

\author{C. Furey}
\affiliation{$ $ \\ Perimeter Institute for Theoretical Physics, Waterloo, Ontario N2L 2Y5, Canada,\\ University of Waterloo, Ontario N2L 3G1, Canada, \\and \\
M.I.T., Cambridge, MA 02139, U.S.A.}

\begin{abstract}




\noindent Unified field theories act to merge the internal symmetries of the standard model into a single group.  Here we lay out something different.  That is, instead of aiming to unify the internal symmetries, we demonstrate a sense in which the group transformations may be unified with the quarks and leptons that they act on.  Similarly, the (3+1) Lorentz transformations may be united with the scalars, spinors, four-vectors and field strength tensors that they act on.  These simplifications occur because the representations can be found in the form of an algebra acting \it on itself. \rm The approach described in this paper is meant to tie everything into the Dixon algebra: $\mathbb{R}\otimes\mathbb{C}\otimes\mathbb{H}\otimes\mathbb{O}$, the tensor product of the only four normed division algebras over $\mathbb{R}$.  Here we demonstrate that the standard model's Lorentz representations may be cast as a special set of \it generalized ideals \rm within the algebra $\CH$. We then make an early attempt at extending this idea to one generation of quarks and leptons.

\end{abstract}

\maketitle

\noindent \bf Introduction. \rm The division algebras are by no means new to physics; most theory, both classical and quantum, is described already in terms of the real, $\mathbb{R}$,  and complex numbers, $\mathbb{C}$.  Furthermore, the Lie group $SU(2)$ is ubiquitous, and its connection to the  quaternions is well known.  It is nearly irresistible to ask if the octonions, $\mathbb{O}$, the last of the set of four normed division algebras over $\mathbb{R}$, have a calling in nature.  Certainly several have thought so: \citep{dixon}-\citep{okubo}, but for the most part, the octonions have remained as a well kept secret from mainstream physics.

More often than not, the octonions are passed by in haste because they are non-associative, and hence at times temperamental.  As we will show, this property is in fact a gift, which will offer a way to streamline some of the standard model's complex structure.

In 1973, G\"{u}naydin and G\"{u}rsey \citep{GGquarks} showed quark structure in the split octonions.  There, they studied the action of the octonionic automorphism group, $G_2$, which contains $SU(3)$ as a subgroup.  In 1994, Dixon published~\citep{dixon}, and using G\"{u}naydin and G\"{u}rsey's model, was the first author to advocate for the use of (two copies of) the full algebra $\mathbb{R}\otimes\mathbb{C}\otimes\mathbb{H}\otimes\mathbb{O}$.     A more detailed comparison between this model and earlier work appears before the conclusion of this text.


The main idea in this current article is to find group representations from the standard model, written in terms of what we will call \it generalized ideals \rm   of the algebra $\RCHO$.  So far this has  uncovered spin 0, spin 1/2 and spin 1 Lorentz representations.  We then propose an extension to a generation of quarks and leptons.
\begin{figure}[h!!]
\includegraphics[width=8.5cm]{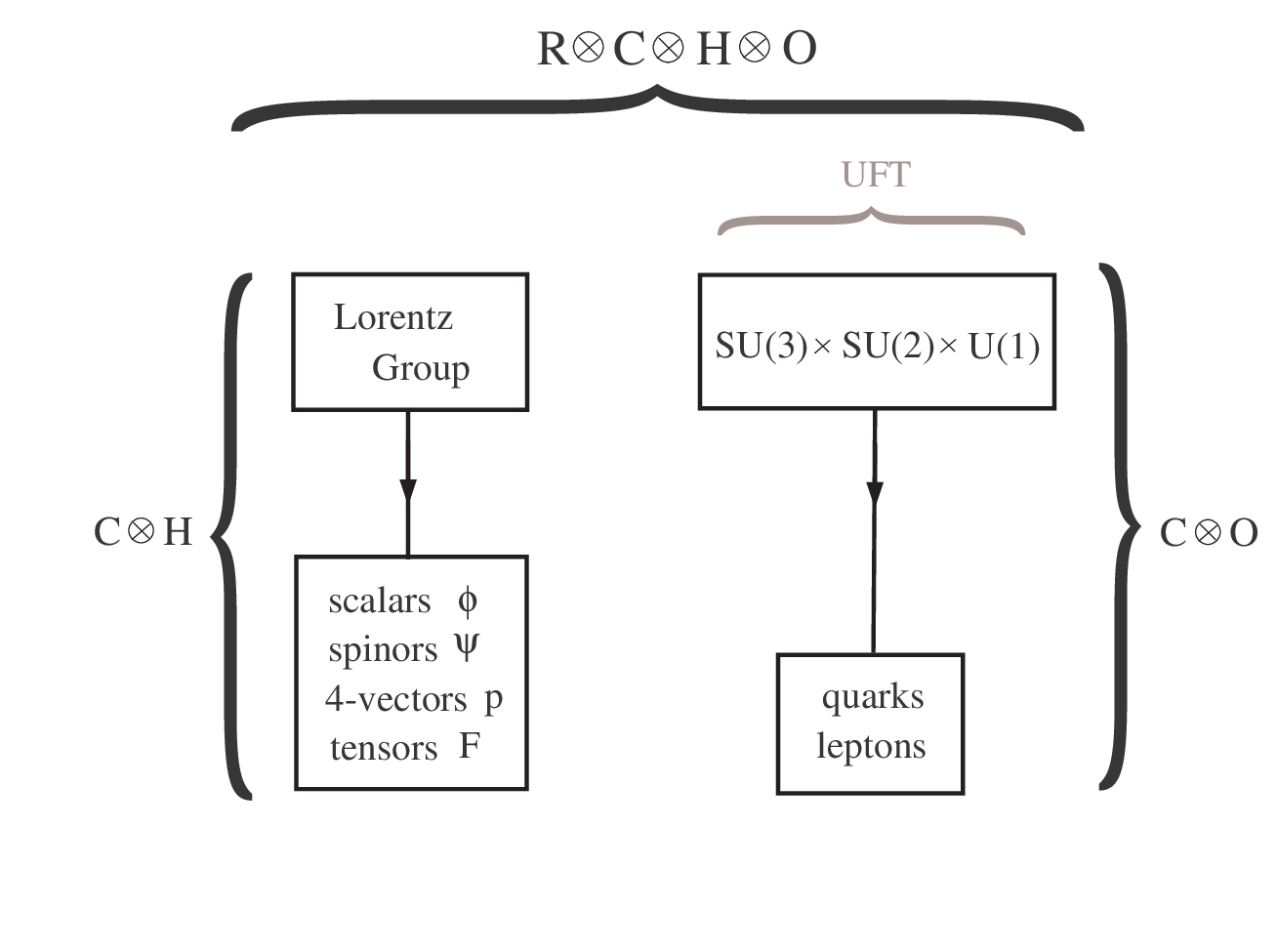}\caption{\label{bigpic}Overview.  Scalars, spinors, 4-vectors and field strength tensors are united with the Lorentz transformations that act on them, via the algebra $\mathbb{C}\otimes\mathbb{H}$.   In like manner,  the internal degrees of freedom for a generation of quarks and leptons, and the generators of $SU(3)\times SU(2)\times U(1)$ can each be represented using the algebra $\mathbb{C}\otimes\mathbb{O}$.  (Note, though, that representations of electroweak generators will require projectors in $\mathbb{C}\otimes\mathbb{H}$ so as to account for chirality.)  Finally,   $\mathbb{C}\otimes\mathbb{H}$ and $\mathbb{C}\otimes\mathbb{O}$ combine, via a tensor product over $\mathbb{C}$, to form $\mathbb{C}\otimes\mathbb{H}\otimes\mathbb{O} = \RCHO$, the tensor product of the only four normed division algebras over the real numbers.  
}
\end{figure}

This work represents the first stage of a larger project.  Towards the end of the text, we will demonstrate how one is able to write down  $SU(3)\times SU(2)\times U(1)$ generators entirely in terms of $\mathbb{R}\otimes\mathbb{C}\otimes\mathbb{H}\otimes\mathbb{O}$, by drawing on the octonions' non-associativity.  However, the internal generators given here do not come from an underlying principle, and so we present them as a provisional solution, until a method is found which is more aesthetically compelling.

\noindent \bf Prerequisites.  \rm  Little algebraic background is needed to understand the following pages, so we provide it here.  Note that all tensor products will be assumed to be over $\mathbb{R}$, unless otherwise stated.

The generic element of $ \mathbb{C} \otimes \mathbb{H}$ is written $a+b\mathbf{i}+c\mathbf{j}+d\mathbf{k}$ where $a,b,c,d \in \mathbb{C}$. $\mathbf{i},\mathbf{j},\mathbf{k}$ follow the non-commutative quaternionic multiplication rules
\begin{equation}
\mathbf{i}^2=\mathbf{j}^2=\mathbf{k}^2=\mathbf{i}\mathbf{j}\mathbf{k}=-1,
\end{equation}
\noindent from which we get $\mathbf{i} \mathbf{j} = -\mathbf{j}\mathbf{i} =  \mathbf{k}, \hspace{0.1cm}  \mathbf{j} \mathbf{k} = -\mathbf{k}\mathbf{j} =  \mathbf{i}, \hspace{0.1cm}  \mathbf{k}\mathbf{i} = - \mathbf{i}\mathbf{k} = \mathbf{j}.$

\begin{figure}[h!]
\includegraphics[width=6cm]{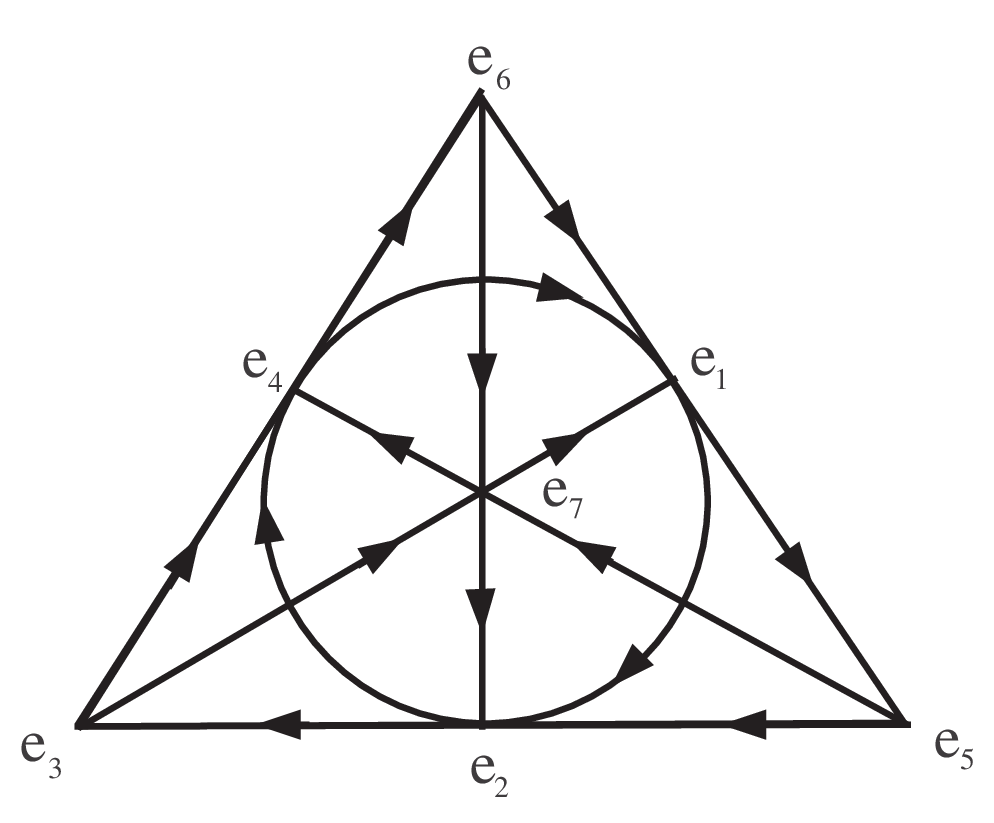}
\caption{\label{fano}
Octonionic multiplication rules}
\end{figure}

The generic element of $\mathbb{C} \otimes \mathbb{O}$ is written $ \sum_{n=0}^7 A_n e_n $, with the $A_n \in \mathbb{C}$.  The $e_n$ are octonionic imaginary units $\left(e_n^2=-1\right)$, apart from $e_0=1$, which multiply according to Figure~\ref{fano}.
Any three imaginary units on a directed line segment in Figure~\ref{fano} act as if they were a quaternionic triple.  For example, $e_6e_1 =-e_1e_6= e_5,$ $e_1e_5=-e_5e_1=e_6,$ $e_5e_6=-e_6e_5=e_1,$ $e_4e_1=-e_1e_4=e_2$, etc.  Octonionic multiplication harbours various symmetries, such as index doubling symmetry:  $e_ie_j=e_k \Rightarrow e_{2i}e_{2j}=e_{2k}$, which can be seen by rotating Figure~\ref{fano} by 120 degrees~\citep{baez}.  For a more thorough introduction to $\mathbb{O}$, see~\citep{baez} and~\citep{okubo}.

The generic element of the Dixon algebra $\mathbb{C} \otimes \mathbb{H} \otimes \mathbb{O}$ is $ \sum_{n=0}^7 B_n e_n $, where the $B_n \in \mathbb{C}\otimes\mathbb{H}$.  Imaginary units of the different division algebras always commute with each other; explicitly, the complex $i$ commutes with the quaternionic $\mathbf{i}, \mathbf{j}, \mathbf{k}$, all four of which commute with the octonionic $\left\{ e_n \right\}$.
\smallskip

We define a subalgebra $I$ of an algebra $A$ to be a \it (generalized) ideal \rm if $m \left( a,\underline{v} \right) \in I,  \forall \underline{v} \in I$ and for any $a \in A$, where $m$ is (generalized) multiplication.  Directly from the definition it can be seen that an ideal can be thought of as an algebra's resilient subspace, which persists no matter which $a$ is multiplied onto it \footnote{As the ideal will pull any element into itself, it can be thought of, loosely speaking, as an algebra's version of a black hole. These ideas originate from a (slowly moving) quantum gravity project, of which the present paper is a byproduct.  The proposal is to identify matter with a web of algebraic expressions.  Given that algebraic expressions supply their own causal structure, they might be able to exist self-sufficiently, without the need for an underlying spacetime. }.

In what is to follow, we will find the invariant subspaces of $\mathbb{C}\otimes\mathbb{H}$ under three separate actions of the algebra on itself:  $m \left( a,\underline{v} \right) = a\underline{v}P + a^*\underline{v}P^*$, $\hspace{0.1cm}$ $m \left( a,\underline{v} \right) = a\underline{v}a^{\dagger}$ and $m \left( a,\underline{v} \right) = a\underline{v}\widetilde{a}$.  Here, $P$ is a projector to be defined shortly. The symbol $*$ denotes the complex conjugate $i \mapsto -i$, the symbol $\sim$ denotes the parity conjugate $\mathbf{i}\mapsto -\mathbf{i},$ $\mathbf{j}\mapsto -\mathbf{j},$ $\mathbf{k}\mapsto -\mathbf{k}$, and the symbol $\dagger$ performs both of these conjugates simultaneously: $i \mapsto -i$, $\mathbf{i}\mapsto -\mathbf{i},$ $\mathbf{j}\mapsto -\mathbf{j},$ $\mathbf{k}\mapsto -\mathbf{k}$.  As with the usual hermitian conjugate of matrices, the order of multiplication is reversed when pairs of algebraic elements are acted upon by $\sim$ or $\dagger$.  For example, $(xy)^{\dagger} = y^{\dagger}x^{\dagger}$.


\noindent \bf Generalized ideals in $\mathbb{C} \otimes \mathbb{H}$.  \rm  It has been known since at least the 1930s, \citep{conway}, that the complex quaternions can encode the Lorentz representations of spinors, four-vectors and the field strength tensor.  This enables a sense of unification in two respects.  First, the Lorentz group actions are now seen to be combined with the vector spaces that they transform.  Furthermore, we see that scalars, spinors, four-vectors, and field strength tensors are each born from the same algebra.  We now go on to show how  these Lorentz representations arise as a special set of generalized ideals within the algebra.

We start with ideals under $m \left( a,\underline{v} \right) = a\underline{v}P + a^*\underline{v}P^*$, from which all three of Weyl, Dirac, and Majorana spinors will arise:
\begin{equation}
\framebox{
${\underline{v}}' =  a\underline{v}P + a^*\underline{v}P^*  $
}\hspace{0.05cm},
\end{equation}
\noindent where $P$ is taken to be $\frac{1+i\mathbf{k}}{2}$.  It should be emphasized, though, that a continuum of choices is available for $P$.

Any element of $\mathbb{C}\otimes\mathbb{H}$ can be written $c_1\frac{1+i\mathbf{k}}{2} + c_2\frac{1-i\mathbf{k}}{2} + c_3\frac{\mathbf{j}+i\mathbf{i}}{2} + c_4\frac{-\mathbf{j}+i\mathbf{i}}{2}$ where the $c_n \in \mathbb{C}$.  It is straightforward to confirm that ${\underline{v}}_1 \equiv   c_1\frac{1+i\mathbf{k}}{2} + c_3\frac{\mathbf{j}+i\mathbf{i}}{2}$ and ${\underline{v}}_2 \equiv c_2\frac{1-i\mathbf{k}}{2} + c_4\frac{-\mathbf{j}+i\mathbf{i}}{2}$ are stable against multiplication by \it any \rm $a \in \mathbb{C}\otimes\mathbb{H}$:
\begin{equation}\label{spinors}
\begin{array}{l}  m(a,  {\underline{v}}_1) = a{\underline{v}}_1P+ a^*{\underline{v}}_1P^* =  a{\underline{v}}_1 \vspace{0.25cm}\\ \hspace{0.2cm} = a \left( c_1\frac{1+i\mathbf{k}}{2} + c_3\frac{\mathbf{j}+i\mathbf{i}}{2} \right) = c_1 ' \frac{1+i\mathbf{k}}{2} + c_3 ' \frac{\mathbf{j}+i\mathbf{i}}{2} \equiv {\underline{v}}_1 ',  \vspace{0.25cm} \hspace{1.5cm}\\  m(a,  {\underline{v}}_2) = a{\underline{v}}_2P+ a^*{\underline{v}}_2P^* =  a^*{\underline{v}}_2 \vspace{0.25cm}\\ \hspace{0.2cm} = a^* \left(c_2\frac{1-i\mathbf{k}}{2} + c_4\frac{-\mathbf{j}+i\mathbf{i}}{2} \right)  = c_2 ' \frac{1-i\mathbf{k}}{2} + c_4 ' \frac{-\mathbf{j}+i\mathbf{i}}{2} \equiv {\underline{v}}_2 ' \end{array}
\end{equation} \noindent for some $c_n' \in \mathbb{C}$, using ${\underline{v}}_1P={\underline{v}}_1$, ${\underline{v}}_1P^*=0$, ${\underline{v}}_2P=0$, ${\underline{v}}_2P^*={\underline{v}}_2\vspace{0.5mm}$.  So we see that $\mathbb{C}\otimes\mathbb{H}$ splits into two ideals under this action.

Now, it is well known that the set $S \equiv \{ \frac{\sigma_x}{2} = \frac{i\mathbf{i}}{2}$, $\frac{\sigma_y}{2} = \frac{i\mathbf{j}}{2}$, $\frac{\sigma_z}{2} = \frac{i\mathbf{k}}{2}$, $\frac{\mathbf{i}}{2}$, $\frac{\mathbf{j}}{2}$, $\frac{\mathbf{k}}{2}\}$ generates the Lorentz algebra, as does $-S^* \equiv \{ \frac{i\mathbf{i}}{2}$, $\frac{i\mathbf{j}}{2}$, $\frac{i\mathbf{k}}{2}$, $-\frac{\mathbf{i}}{2}$, $-\frac{\mathbf{j}}{2}$, $-\frac{\mathbf{k}}{2} \}$.  A linear combination of the generators in $S$, call it ${s}$,  can be exponentiated to form group elements in $ \mathbb{C}\otimes \mathbb{H}$.  According to equations~(\ref{spinors}), we can then certainly write
\begin{equation} \begin{array}{ccc}
e^{i{s}}{\underline{v}}_1 = {\underline{v}}_1 ' & \hspace{0.8cm}  & e^{-i{s}^*}{\underline{v}}_2 = {\underline{v}}_2 '.
\end{array}
\end{equation}


\noindent Much of the above can also be found in~\citep{hes} in terms of Clifford algebras.

Next, let us do some renaming of coefficients:  $c_1 \mapsto \psi^{\uparrow}_{\textup{L}}$, $c_2 \mapsto \psi^{\downarrow}_{\textup{R}}$, $c_3 \mapsto \psi^{\downarrow}_{\textup{L}}$, $c_4 \mapsto \psi^{\uparrow}_{\textup{R}}$, and of basis vectors: $\frac{1+i\mathbf{k}}{2} \mapsto \left[ \uparrow \textup{L} \right]$,
$\frac{1-i\mathbf{k}}{2} \mapsto \left[ \downarrow \textup{R} \right]$, $\frac{\mathbf{j}+i\mathbf{i}}{2} \mapsto \left[ \downarrow \textup{L} \right]$, $\frac{-\mathbf{j}+i\mathbf{i}}{2} \mapsto \left[ \uparrow \textup{R} \right]$.  The reader is encouraged to check that the ideal element ${\underline{v}}_1 = \psi^{\uparrow}_{\textup{L}}\left[ \uparrow \textup{L} \right] + \psi^{\downarrow}_{\textup{L}}\left[ \downarrow \textup{L} \right] \equiv {\underline{\psi}}_{\textup{L}} $ transforms under $e^{i{s}}$ as a left Weyl spinor, and ${\underline{v}}_2 =\psi^{\uparrow}_{\textup{R}}\left[ \uparrow \textup{R} \right] + \psi^{\downarrow}_{\textup{R}}\left[ \downarrow \textup{R} \right] \equiv {\underline{\psi}}_{\textup{R}}$ transforms under $e^{-i{s}^*}$ as a right Weyl spinor.  The $\psi^{\uparrow}$, $\psi^{\downarrow}$ coefficients here are precisely those we are accustomed to writing down in 2-component column vectors.

Combining these ideals, we see that Dirac spinors, given by ${\underline{\psi}}_{\textup{L}}+{\underline{\psi}}_{\textup{R}}$, are then represented by the whole algebra of $\mathbb{C}\otimes\mathbb{H}$.

Furthermore, making use of this multiplication rule yet again, one can show  that $\mathbb{C}\otimes\mathbb{H}$ breaks once more into two different ideals:  (1) the real quaternions, $\mathbb{H}$, and (2) the space obtained by multiplying the real quaternions by the complex $i$.  Each of these, as one can check, begets a Majorana spinor.

So in summary, Weyl, Dirac, and Majorana spinors can each be seen to arise as generalized ideals under the same multiplication rule of the complex quaternions.

By multiplying right and left sides by $\sigma_z = i\mathbf{k}$, one finds that $\left[ \uparrow \textup{L} \right]$, $\left[ \downarrow \textup{L} \right]$, $\left[ \uparrow \textup{R} \right]$ and $\left[ \downarrow \textup{R} \right]$ can alternately be interpreted as $|\uparrow \hspace{0.1cm}\rangle \langle \hspace{0.1cm} \uparrow | $, $|\downarrow \hspace{0.1cm} \rangle \langle \hspace{0.1cm} \uparrow |$, $|\uparrow \hspace{0.1cm} \rangle \langle \hspace{0.1cm} \downarrow |$ and $|\downarrow \hspace{0.1cm} \rangle \langle \hspace{0.1cm} \downarrow |$ respectively. At this stage, then, we see that $\textup{L}$ and $\textup{R}$ are not rigid fixtures, but instead can be rotated freely into each other under right multiplication.  This feature could offer a new perspective on the description of mass.

One feature of this division algebraic framework is that it makes for more streamlined calculations.  As an example, we now introduce a more succint way to conjugate Weyl spinors.


When Weyl spinors are presented in various textbooks, e.g. \citep{bm},  it is sometimes noted that the ``complex conjugate" of the left-handed column spinor $\Phi_{\textup{L}}$ transforms as a right-handed spinor, and vice-versa.  However, the ``complex conjugate" is not really the complex conjugate.  The matrix $\epsilon = -i\sigma_y=\left( \begin{smallmatrix}0&-1\\1&0 \end{smallmatrix}  \right)$ is introduced to ensure that conjugation behaves as it should, and as a result, the ``complex conjugate" is defined as $\epsilon \hspace{.7mm}\Phi_{\textup{L}}^*$.  Furthermore, in order to return back to $\Phi_{\textup{L}}$, one must import an extra factor of $-1$ so that $\Phi_{\textup{L}} = -\epsilon \left( \epsilon  \hspace{.7mm}\Phi_{\textup{L}}^* \right)^*$, instead of just taking the ``complex conjugate" twice.

In our current formalism, however, the true complex conjugate $i\mapsto -i$ makes the above procedure obsolete.  ${\underline{\psi}}_{\textup{L}} ^* = \psi^{\uparrow *}_{\textup{L}}\left[ \uparrow \textup{L} \right]^* + \psi^{\downarrow *}_{\textup{L}}\left[ \downarrow \textup{L} \right]^* = \psi^{\uparrow *}_{\textup{L}}\left[ \downarrow \textup{R} \right] - \psi^{\downarrow *}_{\textup{L}}\left[ \uparrow \textup{R} \right]$, giving the same result as above.

For those with a fondness for metric tensors, it can alternately be said that $i \mapsto -i$ encodes how spinor indices are raised and lowered.
\smallskip

A new multiplication rule leads to the 4-vector representation:
\begin{equation}
\framebox{
${\underline{v}} ' = a \underline{v} a ^{\dagger} $
}\hspace{0.05cm}.
\end{equation}
Any element of $\mathbb{C}\otimes\mathbb{H}$ can be written as a sum of hermitian $\underline{h}=h_0 + h_1i\mathbf{i} +h_2i\mathbf{j} + h_3i\mathbf{k}$ and anti-hermitian $\underline{\bar{h}}=ih_4 + h_5\mathbf{i} +h_6\mathbf{j} + h_7\mathbf{k}$ parts, where the $h_n \in \mathbb{R}$, and the hermitian conjugate is defined as $a^{\dagger} \equiv \widetilde{a}^*$.  
As $a \underline{h} a ^{\dagger}$ is hermitian and $a \underline{\bar{h}} a ^{\dagger}$ is antihermitian for any $a \in \mathbb{C}\otimes\mathbb{H}$, it is clear that $\mathbb{C}\otimes\mathbb{H}$ splits again into two ideals, this time under the multiplication $m\left( a,\underline{v} \right) = a \underline{v} a ^{\dagger}$.

In \citep{gs} it is shown that we may also make use of the complex conjugate of h.  These vector spaces then transform as
\begin{equation} e^{i{s}}\underline{h} e^{-i{s}^{\dagger}}={\underline{h} }'  \hspace{1cm}   e^{-i{s}^*}\underline{h}^* e^{i\widetilde{{s}}}= {\underline{h}^*}',\end{equation}

\noindent where ${s}$ is the same as before.  The antihermitian case follows analogously.
Matching components, one finds that $\underline{h}$ transforms as a contravariant four-vector, and $\underline{h}^*$ as a covariant four-vector.  For example, momentum $p = p_0 + p_1i\mathbf{i} +p_2i\mathbf{j} + p_3i\mathbf{k}$ under a rotation about $\mathbf{k}$ by an angle $\theta$ is given by
\begin{equation} \begin{array}{c}
p' =  e^{-\frac{\theta\mathbf{k}}{2}}p e^{\frac{\theta\mathbf{k}}{2}} = \left(  \cos{\frac{\theta}{2}} - \mathbf{k}\sin{\frac{\theta}{2}}    \right) p \left(    \cos{ \frac{ \theta }{2} } + \mathbf{k}\sin{\frac{\theta}{2}}   \right)=
\\ \\
p_0 + \left( p_1 \cos{\theta} + p_2 \sin{\theta}  \right) i\mathbf{i} + \left(  p_2 \cos{\theta} - p_1 \sin{\theta} \right)i\mathbf{j} + p_3i\mathbf{k},
\end{array}\end{equation}
\noindent as expected.  As shown in~\cite{gs}, the scalar $\frac{1}{2}\left( pq + \widetilde{pq} \right)$ can now form between a covariant vector $p$ and contravariant vector $q$, which is simply the real part of $pq$.  Indeed, when $q=p^*$, this gives $p_0^2 -p_1^2-p_2^2-p_3^2$.

Finally, scalars and field strength tensors can be shown to come from the multiplication rule
\begin{equation}
\framebox{
${\underline{v}} ' = a \underline{v} \widetilde{a}$
}\hspace{0.05cm}.
\end{equation}
$\mathbb{C}\otimes\mathbb{H}$ can be split yet again into generalized  ideals of the form $\underline{\phi} \in \mathbb{C} $ and $\underline{\textup{F}} = \left( F^{32} +iF^{01}\right)\mathbf{i} + \left( F^{13}+iF^{02} \right)\mathbf{j}  + \left( F^{21}+iF^{03}  \right)\mathbf{k}  $, $F^{mn} \in \mathbb{R}$, which weather the multiplication $a \underline{v} \widetilde{a}$ from any element of the algebra.  As ${\underline{\phi} }' = e^{i{s}}\underline{\phi} e^{i\widetilde{{s}}}=\underline{\phi}$, we see that $\underline{\phi}$ transforms as a Lorentz scalar.   In~\citep{hes} and~\citep{gs} it is shown that indeed, massless spin-one bosons are represented by $\underline{\textup{F}}$.  Under the Lorentz transformations,
\begin{equation}e^{i{s}}\underline{\textup{F}} e^{i\widetilde{{s}}}={\underline{\textup{F}}} ' \hspace{1cm}  e^{-i{s}^*}\underline{\textup{F}}^* e^{-i{s}^{\dagger}}={\underline{\textup{F}} ^*} '. \end{equation}

$\underline{\textup{F}} ^* = \left( B^1 +iE^1\right)\mathbf{i} + \left( B^2+iE^2 \right)\mathbf{j}  + \left( B^3+iE^3  \right)\mathbf{k} $ gives the familiar field strength $F_{\mu\nu}$, while $\underline{\textup{F}} $ gives $F^{\mu\nu}$.

\begin{figure}[h!!]
\includegraphics[width=7cm]{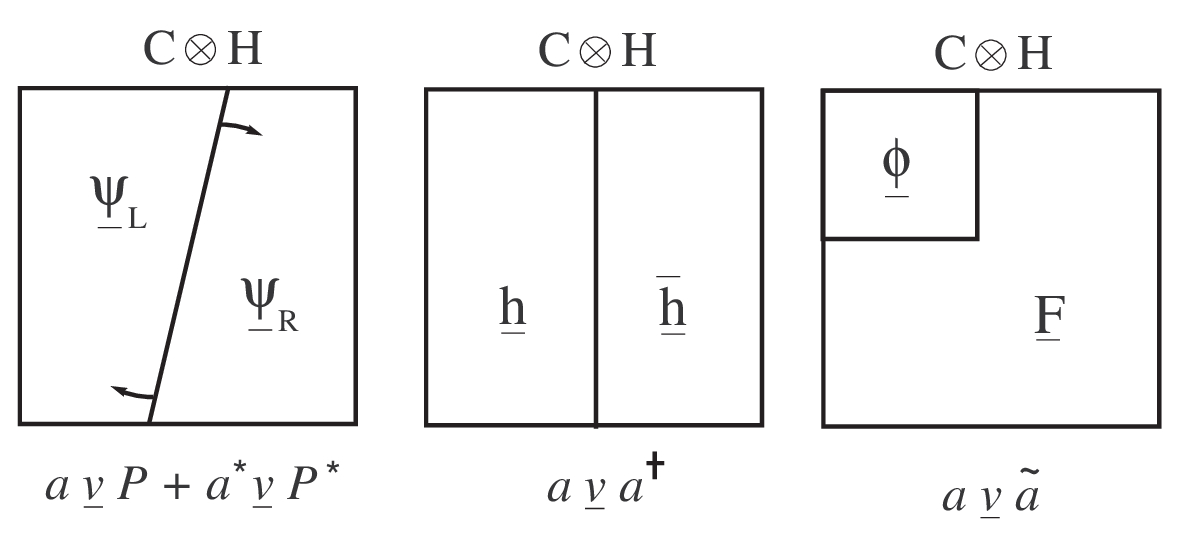}
\caption{\label{CH}
$\mathbb{C}\otimes \mathbb{H}$ decomposes into generalized ideals several times, using three different multiplication rules.  $\mathbb{C}\otimes \mathbb{H}$ breaks down into two ideals (left and right Weyl spinors here) under the multiplication $m\left(a, \underline{v} \right)=a\underline{v}P+a^*\underline{v}P^*$.  Under $m\left(a, \underline{v} \right)=a\underline{v}a^{\dagger}$, $\mathbb{C}\otimes \mathbb{H}$ reduces into hermitian and antihermitian parts, representing four-vectors.  Under $m\left(a, \underline{v} \right)=a\underline{v}\widetilde{a}$, $\mathbb{C}\otimes \mathbb{H}$ breaks down into a scalar and a field strength tensor.   In this formalism, all of the local spacetime degrees of freedom of the standard model come from an algebra of only four complex dimensions.}
\end{figure}

As we have now the spin 0, 1/2 and 1 representations of the Lorentz group (summarized in Figure~\ref{CH}), we have accounted for all of the local spacetime degrees of freedom within the standard model.

Bilinears and other scalars can now be built by combining the various representations, whose Lorentz factors $L$ fit together as do the pieces of a puzzle. Noting that $\widetilde{L} = L^{-1}$, we take for example the real part of $\left({\underline{\psi}}_{\textup{L}}^{\dagger} i \partial \hspace{0.05cm} {\underline{\psi}}_{\textup{L}} +  {\underline{\psi}}_{\textup{R}}^{\dagger} i {\partial}^* \hspace{0.05cm} {\underline{\psi}}_{\textup{R}} \right)' = {\underline{\psi}}_{\textup{L}}^{\dagger} L^{\dagger} \hspace{0.1cm} L^* i \partial \widetilde{L} \hspace{0.1cm} {L}  {\underline{\psi}}_{\textup{L}} + {\underline{\psi}}_{\textup{R}}^{\dagger} \widetilde{L} \hspace{0.1cm} L i {\partial}^* L^{\dagger} \hspace{0.1cm} {L}^*  {\underline{\psi}}_{\textup{R}}  = {\underline{\psi}}_{\textup{L}}^{\dagger} i \partial \hspace{0.05cm} {\underline{\psi}}_{\textup{L}} +  {\underline{\psi}}_{\textup{R}}^{\dagger} i {\partial}^* \hspace{0.05cm} {\underline{\psi}}_{\textup{R}},\vspace{0.7mm}$
where $L\equiv e^{is}$ from before.  This scalar is the same as the familiar ${\overline{\Psi}}_D i \hat{\partial} {\Psi}_D = {{\Psi}}^{\dagger}_D  \beta i \gamma^{\alpha} \partial_{\alpha} {\Psi}_D$ from standard QFT.  However, we point out that in this division algebraic formalism, we need not introduce Dirac matrices explicitly.

Finally we point out, perhaps for the first time, that complex conjugation $i \mapsto -i$ has played the role of the flat space metric tensor (raising and lowering indices) for each of the $\underline{\psi}_L$, $\underline{\psi}_R$, $p$ and $\underline{\textup{F}}$ Lorentz representations.

\noindent \bf Ideals in $\mathbb{C}\otimes\mathbb{O}$. \rm  Given that the $\mathbb{C}\otimes\mathbb{H}$ part of $\RCHO$ holds the local spacetime degrees of freedom, we might suspect that the $\mathbb{C}\otimes\mathbb{O}$ part holds the internal degrees of freedom pertaining to $SU(3)\times SU(2)\times U(1)$. 

\noindent \bf Quarks and Leptons. \rm In what follows, we will need to make use of the eigenvalue equations which arise from the action of the algebra on itself.  Let $ \mathcal{O} \left( \underline{v} \right) = c \hspace{0.05cm} \underline{v}$ be the generic form of such an eigenvalue equation where $\mathcal{O}$ is an operator constructed from elements of the algebra $A$, $\underline{v}$ is the solution, which is also in $A$, and $c \in \mathbb{C} \subset A$.

A straightforward example is left multiplication by $\mathcal{O} = \frac{\sigma_z}{2} = \frac{i\mathbf{k}}{2}$ in $\CH$ from the previous section, which has four solutions as opposed to just the usual spin up and down pair.  For $c = \frac{1}{2}$, we have $\underline{v} = \left[ \uparrow \textup{L} \right]$ and $\underline{v} = \left[ \uparrow \textup{R} \right]$.  For $c = -\frac{1}{2}$, we have $\underline{v} = \left[ \downarrow \textup{L} \right]$ and $\underline{v} = \left[ \downarrow \textup{R} \right]$.  

\begin{figure}[h!]
\includegraphics[width=7cm]{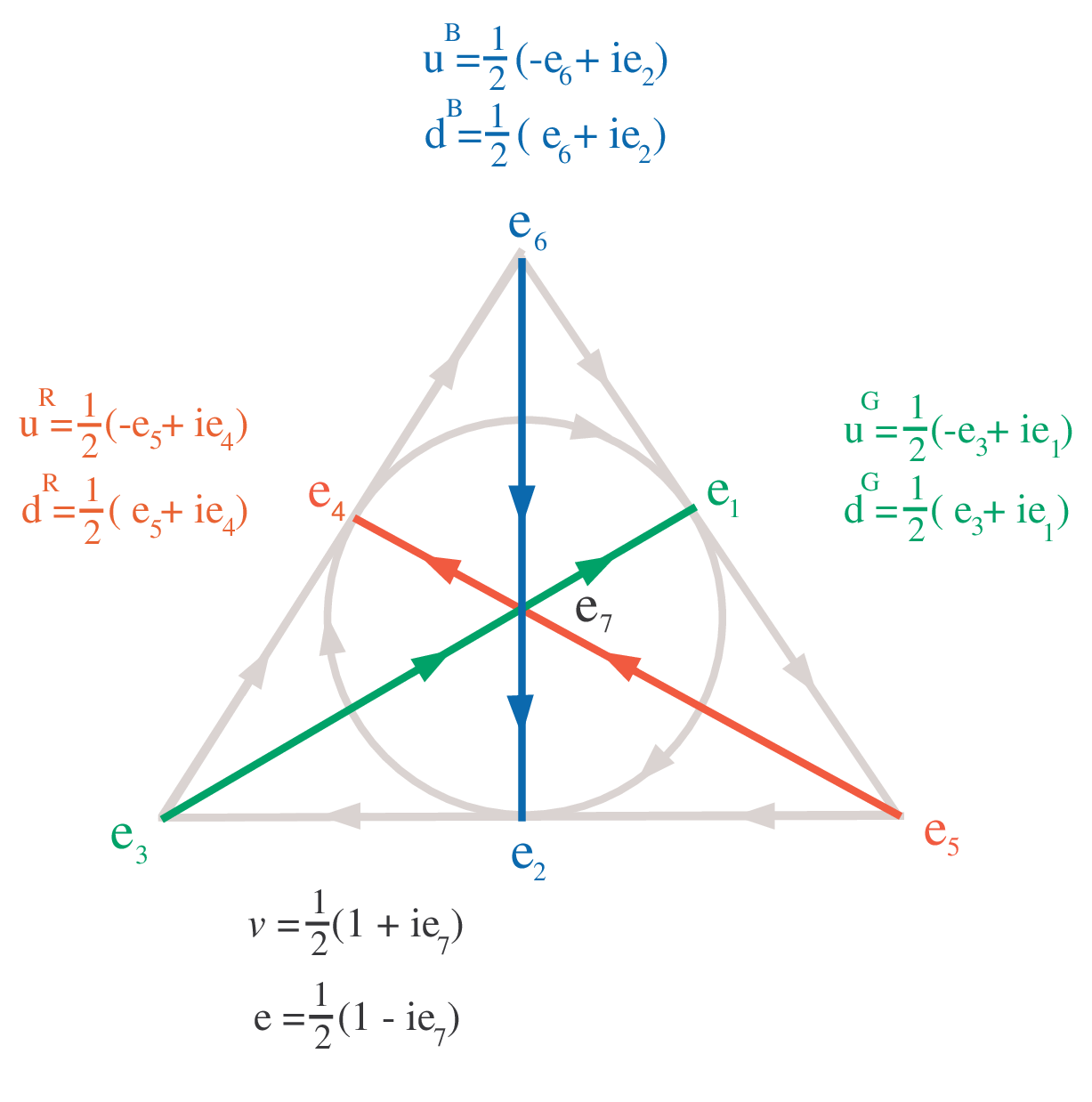}
\caption{\label{fermion}
Eigenvectors of the weak isospin operator, $\frac{ie_7}{2}$, give a full generation of fermions as a basis for $\mathbb{C}\otimes\mathbb{O}$.}
\end{figure}

Replacing the quaternionic imaginary $\mathbf{k}$ with the octonionic imaginary $e_7$ gives another operator $\mathcal{O} = \frac{ie_7}{2}$.  Recall that the octonions can be thought of as a net of quaternionic triples, where each line in Figure~\ref{fano} behaves like $\left\{ \mathbf{i}, \mathbf{j}, \mathbf{k} \right\}$.  Perhaps unsurprisingly, we find eight solutions to $\frac{ie_7}{2} \underline{v} = c \hspace{0.05cm} \underline{v}$, suggestively named here in Figure~\ref{fermion}.  They span all of $\mathbb{C}\otimes\mathbb{O}$, meaning that one generation of fermions trivially constitutes an ideal.


Immediately striking from Figure~\ref{fermion} is the fact that the index doubling symmetry of octonions gives (discrete) colour symmetry of particles.  These rotations of 120 degrees map $u^R \mapsto u^G \mapsto u^B \mapsto u^R$, and $d^R \mapsto d^G \mapsto d^B \mapsto d^R$.  Furthermore, as the leptons are linear combinations of 1 and $e_7$, they are invariant under index doubling, which expresses their immunity to such transformations.

\noindent \bf Provisional Bosons. \rm For concreteness, we would like to write down the twelve linearly independent generators of the standard model gauge group.  As of yet, these objects are not shown to come from some deeper principle;  this is currently an active area of research.  Instead, we write the generators using left multiplication of the algebra on itself, so as to demonstrate how the non-associativity of the octonions makes possible their description.

Let us start by trying to find the generators of $SU(3)$. As the leptons are singlets under $SU\left(3\right)$, we will need the generators to annihilate them, and when acting on quarks, we ask that the generators map one colour to another according to the Gell-Mann matrices.  It is not hard to verify that in fact no number in $\mathbb{C}\otimes\mathbb{O}$ is capable of this.

So it seems that multiplying fermions by a single element of $\mathbb{C}\otimes\mathbb{O}$ is of no help.  However, perhaps unexpectedly, we find that multiplying by a sequence of elements succeeds.

Associativity would have assured that multiplication on a fermion $\underline{f}$ by any two numbers $a_1$ and $a_2$ could be summarized into a single multiplication  $a_1' \underline{f}$:  
\begin{equation}a _2\left( a _1 \underline{f} \right) = \left( a _2 a _1 \right) \underline{f} = a_1' \underline{f},
\end{equation}
\noindent where $a_{1}'=a_{2} a_{1}$.  As luck would have it though, the octonions are non-associative and it is not hard to find examples of sequences of multiplications which can not be summarized.  Take for example the right-to-left  multiplication of $e_3$, $e_4$ on the blue down quark:  $\overleftarrow{e_{34}}\left(\frac{1}{2}\left[ e_6+ie_2 \right]\right) \equiv  e_3 \left( e_4 \left(\frac{1}{2}\left[ e_6+ie_2 \right] \right) \right) = \frac{1}{2} \left[ -1 +ie_7 \right]=-\textup{e}$.  This is not the same as $\left( e_3  e_4 \right) \left(\frac{1}{2}\left[ e_6+ie_2 \right] \right)= \left( e_6 \right) \left(\frac{1}{2}\left[ e_6+ie_2 \right] \right)= \frac{1}{2} \left[ -1 -ie_7 \right]=-\nu$, and in fact there exists no $a \in \mathbb{C}\otimes\mathbb{O}$ such that $\overleftarrow{e_{34}}\left(\frac{1}{2}\left[ e_6+ie_2 \right]\right) = a \left(\frac{1}{2}\left[ e_6+ie_2 \right]\right)$.

Complex linear combinations of the sequences are equivalent to the (associative) set of 8 by 8 complex matrices, as is already provided in references \citep{dixon} and \citep{Leo}.  Crucially, this explains how associative groups arise  from the non-associative octonions.

Translating the appropriate 8 by 8 complex matrices into octonionic chains, we find
\begin{equation}
\begin{array}{ll}
\lambda_{1} = \frac{1}{2}\left(  -\overleftarrow{e_{157}} +\overleftarrow{e_{347}} \right) &
\lambda_{2} = -\frac{i}{2}\left(  \overleftarrow{e_{14}} + \overleftarrow{e_{35}} \right) \vspace{0.25cm}\\
\lambda_{3} = \frac{1}{2}\left(  \overleftarrow{e_{137}} - \overleftarrow{e_{457}} \right) &
\lambda_{4} = -\frac{1}{2}\left(  \overleftarrow{e_{257}} + \overleftarrow{e_{467}} \right) \vspace{0.25cm}\\
\lambda_{5} = \frac{i}{2}\left(  -\overleftarrow{e_{24}} + \overleftarrow{e_{56}} \right) &
\lambda_{6} = -\frac{1}{2}\left(  \overleftarrow{e_{167}} + \overleftarrow{e_{237}} \right)\vspace{0.25cm}\\
\lambda_{7} = \frac{i}{2}\left(  \overleftarrow{e_{12}} + \overleftarrow{e_{36}} \right) &
\lambda_{8} = \frac{1}{2\sqrt{3}}\left(  -\overleftarrow{e_{137}} - \overleftarrow{e_{457}} +2\overleftarrow{e_{267}} \right),
\end{array}\end{equation}
\noindent which annihilate leptons and map quarks according to the Gell-Mann matrices.  These generators obey the analogue of the commutation relations
\begin{equation}\left[ \hspace{0.1cm} \frac{\lambda_{\ell}}{2}, \hspace{0.1cm} \frac{\lambda_{m}}{2}\hspace{0.1cm} \right] \hspace{0.05cm} \underline{f} = ic_{\ell m n}\frac{\lambda_{n}}{2} \underline{f}  \hspace{1cm} \forall \hspace{0.15cm} \underline{f},\end{equation}
\noindent where $c_{\ell m n}$ are the structure constants of $su\left(3\right)$.  


Commuting with this set are $SU(2)$ generators
\begin{equation}
\begin{array}{lll}
\tau_1 =  -\overleftarrow{e_{124}} \left( \hspace{0.3cm} \right)P  \hspace{.3cm}&
\tau_{2} =  -\overleftarrow{e_{356}}\left( \hspace{0.3cm} \right) P \hspace{.3cm}&
\tau_{3}  = i e_7\left( \hspace{0.3cm} \right) P,
\end{array}
\end{equation}
\noindent where right multiplication by $P=\frac{1}{2}\left( 1 + i\mathbf{k} \right)$ projects to keep only the left handed fermions.

In the remaining space, which commutes with both of these Lie algebras, we have a generator acting as weak hypercharge,
\begin{equation}Y = \frac{1}{6} \left( \overleftarrow{e_{137}} +\overleftarrow{e_{267}} + \overleftarrow{e_{457}}   \right)+\frac{i}{2}e_7 \left( \hspace{0.3cm} \right) P^*,
\end{equation}\noindent whose eigenvalues are -1/2 for left-handed leptons, 1/6 for left-handed quarks, -1 for the right-handed electron, 2/3 for right-handed up quarks, -1/3 for right-handed down quarks, and 0 for a \it right-handed neutrino, \rm naturally present in the algebra.


\noindent \bf Antiparticles.\hspace{0.2cm}\rm From our earlier treatment of complex conjugating Weyl spinors, and from noting the form of antiquarks in~\citep{GGquarks}, it is clear that this algebra maps particles to antiparticles using \it only \rm complex conjugation $i \mapsto -i$.  Explicitly, the basis vectors for the antiparticles corresponding to Figure~\ref{fermion} are given by $\nu ^* = \frac{1-ie_7}{2}$, $e^*=\frac{1+ie_7}{2}$, ${u^R}^*=\frac{-e_5-ie_4}{2}$, ${d^R}^*=\frac{e_5-ie_4}{2}$, $\dots$

The familiar conjugates $*$, $\sim$, $\dagger$ and also the octonionic analogues of these, should lead to discrete symmetries of the theory. Care should be taken, though, when matching conjugates to C, P and T.  For example, if one chooses to write a spacetime four-vector in our anti-hermitian ideal as $\mathbf{x} = ti + x\mathbf{i} + y\mathbf{j} + z\mathbf{k}$, then complex conjugation $i\mapsto -i$ is also able to induce time reversal.

\noindent \bf Contrast with Earlier Work. \rm As the results of this paper were found independently of  \citep{dixon} and \citep{GGquarks}, many aspects of the model are different. Here, we identify $\mathbb{C}\otimes\mathbb{O}$ with a full generation of quarks and leptons, instead of half of a generation plus antiparticles, as is found in the earlier literature.  

Furthermore, Figure~\ref{bigpic} shows how our work separates $SU(3)\times SU(2)\times U(1)$ and local spacetime degrees of freedom into the algebras $\mathbb{C}\otimes\mathbb{O}$ and  $\mathbb{C}\otimes\mathbb{H}$ respectively.  In contrast, \citep{dixon} uses $\mathbb{C}\otimes\mathbb{O}$ to describe $SU_C(3)$ and $U_Y(1)$, and $\mathbb{C}\otimes\mathbb{H}$ to describe the (3+1) Lorentzian degrees of freedom, and the internal symmetry $SU_L(2)$. 
Readers are encouraged to study~\citep{dixon}, \citep{GGquarks} and the early division algebraic work of G\"{u}naydin and G\"{u}rsey.


\noindent \bf Current Work.\hspace{0.2cm}\rm  This article shows that  $SU(3)\times SU(2) \times U(1)$ generators can come from the algebra, but as with the standard model, it leaves some mysteries hanging:  Why is $SU(3)\times SU(2) \times U(1)$ nature's preferred group?  Or is it?  Why does $SU(2)$ transform the left and not the right?

The next goal of this project is to find the internal symmetry generators, hopefully coming from some ideal principle, preferably mirroring the forms of the ideals found in $\mathbb{C}\otimes\mathbb{H}$.  If this can be done, it might lend an explanation as to why particles are governed by $SU(3)\times SU(2) \times U(1)$, as opposed to the octonionic automorphism group $G_2$.

\noindent \bf Conclusion. \rm   This article puts forward
the proposal that the group representations of fundamental particles could ultimately come from a single algebra acting on itself.  Specifically, they are proposed to arise from $\RCHO$ in the form of \it generalized ideals. \rm 

\noindent \bf Acknowledgement. \rm 
Thank you to F. Caravelli, G. Dixon, B. Hartmann, J. Koeplinger, T. Konopka, S. Lloyd, P.L. Mana, R. Sorkin and especially G. Fiore for his detailed suggestions.  Thank you to NSERC, the University of Waterloo and the Perimeter Institute.  Research at the Perimeter Institute for Theoretical Physics is supported in part by the Government of Canada through NSERC and by the Province of Ontario through MRI.

\end{document}